\documentclass[aps,prd,amsmath, amssymb,eqsecnum,
nofootinbib,showpacs,tightenlines]{revtex4}
\global\arraycolsep=2pt
\usepackage{bm}
\usepackage{epsfig}

\newcommand{\ben}{\begin{equation*}}
\newcommand{\een}{\end{equation*}}
\newcommand{\bean}{\begin{eqnarray*}}
\newcommand{\eean}{\end{eqnarray*}}

\newcommand{\nn}{\nonumber}
\newcommand{\be}{\begin{equation}}
\newcommand{\ee}{\end{equation}}
\newcommand{\bea}{\begin{eqnarray}}
\newcommand{\eea}{\end{eqnarray}}
\newcommand{\binomial}[2]{ \begin{pmatrix}#1\\#2\end{pmatrix} }
\begin{document}
\title{Multiple Scattering Methods in Casimir Calculations}

\author{Kimball A. Milton} 
\email{milton@nhn.ou.edu}
\homepage{http://www.nhn.ou.edu/%7Emilton}

\author{Jef Wagner}
\email{wagner@nhn.ou.edu}
\affiliation{Oklahoma Center for High Energy Physics 
and Homer L. Dodge Department of Physics and Astronomy,
University of Oklahoma, Norman, OK 73019, USA}
\date{\today}
\pacs{03.70.+k, 03.65.Nk, 11.80.Et, 11.80.La}

\begin{abstract}
Multiple scattering formulations have been %recently rediscovered 
employed for more than 30 years as
a method of studying the quantum vacuum or Casimir interactions between
distinct bodies.  %The methods are hardly new, but increased computing
%power and advances in understanding
% allow us to extract information efficiently.  
Here we review the
method in the simple context of $\delta$-function potentials, so-called
semitransparent bodies. (In the limit of strong coupling, a semitransparent
boundary becomes a Dirichlet one.) After applying the method to rederive the
Casimir force between two semitransparent plates and the Casimir self-stress
on a semitransparent sphere, we obtain expressions for the Casimir
energies between disjoint parallel semitransparent cylinders 
and between disjoint 
semitransparent spheres.  Simplifications occur for weak and strong coupling.
In particular, after performing a power series expansion in the ratio of
the radii of the objects to the separation between them, we are
able to sum the weak-coupling expansions exactly to obtain explicit
closed forms for the Casimir interaction energy.  The same can be done
for the interaction of a weak-coupling sphere or cylinder with a Dirichlet
plane. We show that the proximity force approximation (PFA), which
becomes the proximity force theorem when the objects are almost touching, 
is very poor for finite separations.  
\end{abstract}

\maketitle
\section{Introduction}
Recently, there has been a flurry of papers concerning ``exact'' methods of
calculating Casimir energies or forces between arbitrary distinct bodies.
Most notable is the recent paper by Emig, Graham, Jaffe, and Kardar 
\cite{Emig:2007cf}. (Details, applied to a scalar field, are supplied
in Ref.~\cite{Emig:2007me}. See also Refs.~\cite{Emig:2007qw,sidewalls}.) 
%In fact it is clear that the
%methods are not so novel: Certainly the multiple scattering method was 
%explicit in the 
Precursors include an early paper of Renne \cite{renne},
rederiving the Lifshitz formula \cite{lifshitz} in this way,
the famous papers of Balian and Duplantier 
\cite{Balian:1977qr,Balian:1976za,Balian:2004jv},
%Multipole expansion
%methods can be traced back at least as far as Sommerfeld \cite{Sommerfeld}. 
%Most explicitly, an earlier drafted paper by 
work of Kenneth and Klich 
\cite{Kenneth:2007jk} based on
%appeared which shows that
% the basis of the approach lies in 
the Lippmann-Schwinger formulation
of scattering theory \cite{lippmann},  %In fact, the most related
%precursors seem to be 
 the papers by Bulgac, Marierski, and Wirzba 
\cite{wirzba07,Bulgac:2005ku,Wirzba:2005zn}, who use the
 modified Krein formula \cite{krein},
and by Bordag \cite{Bordag:2006vc,Bordag:2006kx}, who derives his
results from a path integral formulation.
Dalvit et al.\  \cite{Mazzitelli:2006ne, Dalvit:2006wy}
use the argument principle to calculate the interaction between
conducting cylinders %(of length $L$) 
with parallel axes.
See also Reynaud et al.~\cite{reynaud} and references therein.

In fact, Emig and earlier collaborators 
\cite{buscher,Emig:2006uh,Emig:2002xz}
have published a series of papers, using closely related methods
to calculate numerically forces between distinct bodies,
starting from periodically deformed ones.
%They start from the change in the density of states,
%\begin{subequations}
%\bea
%E&=&\frac{\hbar c}2\int_0^\infty dq\,q\,\delta \rho(q),\\
%\delta\rho(q)&=&-\frac1\pi\frac\partial{\partial q}\mbox{Tr}
%\ln\mathcal{MM}_\infty^{-1},\eea
%\end{subequations}
%where the matrix operator $\mathcal{M}$ is given by the Euclidean
%Green's function
%\be
%G_0(\mathbf{x,x'},q)=\frac1{4\pi|\mathbf{x-x'}|}e^{-q|\mathbf{x-x'}|}\ee
%evaluated on the (Dirichlet, for example) surfaces. $\mathcal{M}_\infty$
%is defined at infinite surface separation.
%Emig first used this method to calculate the force between corrugated
%surfaces \cite{Emig:2002xz}.
%They later used the technique to calculate the exact force between a
%cylinder (radius $a$) and a plate (distance of closest approach $d$)
%\cite{Emig:2006uh}.  The determinant is obtained by a 
%truncation on partial waves; $l=25$ is sufficient even for $d/a=0.1$.  
Strong deviation from the proximity force approximation (PFA) 
is seen for when the distance between the bodies
is large compared to their radii of curvature.  
Bordag
\cite{Bordag:2006vc}
has precisely quantified the first correction to the PFA both for a cylinder
and a sphere near a plane.
As Gies and Klingm\"uller note
\cite{Gies:2006bt},
1\% deviations from the PFA occur when the ratio of 
the distance between the cylinder and the plate to the radius of the
cylinder exceeds
$0.01$.  We will not discuss the worldline method of Gies and collaborators
\cite{Gies:2006xe,Gies:2006cq,Gies:2003cv}
further, as that method lies rather outside our discussion here.  
Similar remarks apply to the work of Capasso et al.~\cite{capasso}, who
calculate forces from stress tensors using the familiar construction
of the stress tensor in terms of Green's
dyadics \cite{Schwinger:1977pa,Milton:1978sf}. They use a numerical 
finite-difference engineering method.% Finite-difference
%frequency-domain methods are employed in two dimensions to obtain forces
%between metal squares and plates to 3\% accuracy,  ``using reasonable
%computational resources.'' There may be problems with scalability of
%this procedure.

It is clear, then, with the exception of these last two methods, these
approaches are fundamentally equivalent.  We will refer to all of
the former methods as multiple scattering techniques.  We will now
proceed to state the formulation in a simple, straightforward
way, and apply it to various situations, all characterized by $\delta$-function
potentials. (A preliminary version of some of our results has already
appeared \cite{Milton:2007gy}.)

\section{Formalism}
We begin by noting that %as others have observed, the derivation of
%the chief result of Emig et al.\ \cite{Emig:2007cf} is much more general than 
%that given in their paper.  In fact, it is a consequence of 
the multiple-scattering formalism may be derived from the general formula
for Casimir energies (for simplicity here we restrict attention to a
massless scalar field) \cite{Schwinger75}
\be
E=\frac{i}{2\tau}\mbox{Tr}\ln G,\label{trln}
\ee
where $\tau$ is the ``infinite'' time that the
configuration exists, and $G$ is the Green's function in the
presence of a potential $V$ satisfying 
(matrix notation)
\be
(-\partial^2+V)G=1, \ee
subject to some boundary conditions at infinity.  (For example,
we can use causal or Feynman boundary conditions, or, alternatively,
retarded Green's functions.)
In Appendix \ref{Appa} we give a heuristic derivation of this
fundamental formula.

The above formula for the Casimir energy is defined up to an infinite
constant, which can be at least partially compensated by inserting
a factor as do Kenneth and Klich \cite{Kenneth:2007jk}:
\be E=\frac{i}{2\tau} \mbox{Tr}\ln G G_0^{-1}.\ee
Here $G_0$ satisfies, with the same boundary conditions as $G$, the
free equation
\be -\partial^2 G_0=1.\ee

Now we define the $T$-matrix (note that our definition of $T$ differs by
a factor of 2 from that in Ref.~\cite{Emig:2007cf})
\be
T=S-1=V(1+G_0V)^{-1}.\ee
We then follow standard scattering theory \cite{lippmann}, as reviewed
in Kenneth and Klich \cite{Kenneth:2007jk}. 
(Note that there seem to be some sign and ordering errors in that reference.)
The Green's function can be alternatively written as
\be
G=G_0-G_0TG_0
=\frac1{1+G_0V}G_0=V^{-1}TG_0,\ee
which results in two formul\ae\ for the Casimir energy
\begin{subequations}
\bea
E&=&\frac{i}{2\tau}\mbox{Tr}\ln \frac1{1+G_0V}\label{e1}\\
&=&\frac{i}{2\tau}\mbox{Tr}\ln V^{-1}T.\label{e2}\eea
\end{subequations}

If the potential has two disjoint parts,
\be
V=V_1+V_2,\ee
it is easy to show that
\be
T=(V_1+V_2)(1-G_0T_1)(1-G_0T_1G_0T_2)^{-1}(1-G_0T_2),\ee
where
\be
T_i=V_i(1+G_0V_i)^{-1},\quad i=1,2.\ee
Thus, we can write the general expression for the interaction between
the two bodies (potentials) in two alternative forms:
\begin{subequations}
\bea
E_{12}&=& -\frac{i}{2\tau}\mbox{Tr}\ln(1-G_0T_1G_0T_2)\label{gtgt}\\
&=&-\frac{i}{2\tau}\mbox{Tr}\ln(1-V_1G_1V_2G_2),\label{vgvg}
\eea
\end{subequations}
where
\be
G_i=(1+G_0V_i)^{-1}G_0,\quad i=1,2.\ee
The first form is exactly that given by Emig et al.~\cite{Emig:2007cf},
and by Kenneth and Klich \cite{Kenneth:2007jk}, 
while the latter is actually easily used if we know the individual
Green's functions.  (The effort involved in calculating with either is
identical.)  In fact, the general form (\ref{gtgt}) was recognized
earlier and applied to planar geometries by Maia Neto, Lambrecht, and
Reynaud \cite{maianeto,lambrecht,reynaud}.  In fact, Renne \cite{renne}
essentially used Eq.~(\ref{vgvg}) to derive the Lifshitz formula in 1971.

\section{Casimir interaction between $\delta$-plates}
We now use the second formula above (\ref{vgvg}) to calculate the Casimir energy
between two parallel semitransparent plates, with potential 
\be V=\lambda_1 \delta(z-z_1)+\lambda_2\delta (z-z_2),\ee
where the dimension of $\lambda_i$ is $L^{-1}$.
The free reduced Green's function is (where we have performed the
evident Fourier transforms in time and the transverse directions)
\be
g_0(z,z')=\frac1{2\kappa}e^{-\kappa|z-z'|},\quad \kappa^2=\zeta^2+k^2.\ee
Here $\mathbf{k}=\mathbf{k_\perp}$ is the transverse momentum, and 
$\zeta=-i\omega$ is the Euclidean frequency.
The Green's function associated with a single $\delta$-function potential is
\be
g_i(z,z')=\frac1{2\kappa}\left(e^{-\kappa|z-z'|}-\frac{\lambda_i}{\lambda_i
+2\kappa} e^{-\kappa|z-z_i|}e^{-\kappa|z'-z_i|}\right).\ee
Then the energy/area is 
\be
\mathcal{E}=\frac1{16\pi^3}\int d\zeta\int d^2k\int dz\ln(1-A)(z,z),\ee
where, in virtue of the $\delta$-function potentials ($a=|z_2-z_1|$)
\bea
A(z,z')&=&
\frac{\lambda_1\lambda_2}{4\kappa^2}\delta(z-z_1)
\left(1-\frac{\lambda_1}{\lambda_1+2\kappa}\right)
 e^{-\kappa|z_1-z_2|}
\left(1-\frac{\lambda_2}{\lambda_2+2\kappa}\right)
e^{-\kappa|z'-z_2|}\nn\\
&=&\frac{\lambda_1}{\lambda_1+2\kappa}\frac{\lambda_2}{\lambda_2
+2\kappa}e^{-\kappa a}e^{-\kappa|z'-z_2|}\delta(z-z_1).\eea
We expand the logarithm according to
\be
\ln(1-A)=-\sum_{s=1}^\infty \frac{A^s}s.\label{logexp}\ee
For example, the leading term is easily seen to be
\be
\mathcal{E}^{(2)}=-\frac{\lambda_1\lambda_2}
{16\pi^3}\int \frac{d\zeta \,d^2k}{4\kappa^2}
 \,e^{-2\kappa a}
=-\frac{\lambda_1\lambda_2}{32\pi^2a},\label{wcpp}\ee
which uses the change to polar coordinates,
\be
d\zeta \,d^2k=d\kappa\,\kappa^2\,d\Omega.\ee
In general, it is easy to check that, because $A(z,z')$ factorizes here,
$A(z,z')=B(z)C(z')$, $\mbox{Tr}\,A^n=(\mbox{Tr}\,A)^n$, or
\be
\mbox{Tr}\,\ln(1-A)=\ln(1-\mbox{Tr}\,A),\ee
so the Casimir interaction between the two semitransparent plates is
\be
\mathcal{E}=\frac1{4\pi^2}\int_0^\infty d\kappa \,
\kappa^2\ln\left(1-\frac{\lambda_1}
{\lambda_1+2\kappa}e^{-\kappa a}\frac{\lambda_2}{\lambda_2+2\kappa}
e^{-\kappa a}\right),\ee
which is exactly the well-known result \cite{Milton:2007ar}.

\section{Casimir self-energy for a single semitransparent sphere}
Before we embark on new calculations, let us also confirm the known
result for the self-stress on a single sphere of radius
$a$ using this formalism.  (This demonstrates, as did the rederivation
of the Boyer result \cite{Boyer:1968uf} by Balian and 
Duplantier \cite{Balian:1977qr}, that the multiple scattering
method is equally applicable to the calculation of self-energies.) 
We start from the general formula (\ref{e1}), where
\be
V(\mathbf{r,r'})=\lambda\delta(r-a)\delta(\mathbf{r-r'}).\label{sppot}
\ee
We use the Fourier representation for the propagator in Euclidean space,
\be
G_0(\mathbf{r,r'})=\frac{e^{-|\zeta||\mathbf{r-r'}|}}{4\pi|\mathbf{r-r'}|}
=\int\frac{d^3k}{(2\pi)^3}\frac{e^{i\mathbf{k\cdot(r-r')}}}
{k^2+\zeta^2},\ee
as well as the partial wave expansion of the plane wave
\be
e^{i\mathbf{k\cdot r}}=\sum_{lm}4\pi i^lj_l(kr)Y_{lm}(\mathbf{\hat r})Y_{lm}^*
(\mathbf{\hat k}).\label{pwepw}
\ee
Then, from the orthonormality of the spherical harmonics,
\be
\int d\mathbf{\hat k} Y_{lm}^*(\mathbf{\hat k})Y_{l'm'}(\mathbf{\hat k})=
\delta_{ll'}\delta_{mm'},
\ee
we obtain the representation
\be
G_0(\mathbf{r,r'})=\frac2\pi\sum_{lm}\int_0^\infty \frac{dk\,k^2}{k^2+\zeta^2}
j_l(kr)j_l(kr')Y_{lm}(\mathbf{\hat r})Y_{lm}^*(\mathbf{\hat r'}).\ee

Now we combine the representation for the free Green's function with the 
spherical potential (\ref{sppot}) to obtain
\be
(G_0V)(\mathbf{r,r'})=\frac{2\lambda}\pi\delta(r'-a)\sum_{lm}\int_0^\infty
\frac{dk\,k^2}{k^2+\zeta^2} j_l(ka)j_l(kr)Y_{lm}(\mathbf{\hat r})
Y_{lm}^*(\mathbf{\hat r'}).\label{gv}
\ee
When this, or powers of this, is traced (that is, $\mathbf{r}$ and 
$\mathbf{r'}$
are set equal, and integrated over), we obtain a poorly defined expression;
to regulate this, we assume $r\ne a$, for example, $r<a$.  (This is a type
of point-split regulation.)  Then, because
\be
j_l(ka)=\frac12\left(h_l^{(1)}(ka)+h_l^{(2)}(ka)\right)=
\frac12\left(h_l^{(1)}(ka)+(-1)^lh_l^{(1)}(-ka)\right),
\ee
while $j_l(kr)=(-1)^lj_l(-kr)$, we see that the $k$ integration in 
Eq.~(\ref{gv}) can be evaluated as\footnote{Of course, this result is the
immediate consequence of the usual partial wave expansion
$$G_0(\mathbf{r,r'})=ik\sum_{lm}j_l(kr_<)h^{(1)}_l(kr_>)Y_{lm}(\mathbf{\hat r})
Y^*_{lm}(\mathbf{\hat r'}),\quad k=|\omega|.$$
The point of our slightly more elaborate approach here is that it generalizes to
the corresponding two-body case---see Eq.~(\ref{contourint}).}
\be
\int_0^\infty\frac{dk\,k^2}{k^2+\zeta^2} j_l(ka)j_l(kr)=
\frac\pi{a}K_{l+1/2}(|\zeta|a)I_{l+1/2}(|\zeta| r), \quad r<a.\ee
Thus, it is easily seen that an arbitrary power of $G_0V$ has trace
\be
\mbox{Tr}\,(G_0V)^n=(\lambda a)^n\sum_{lm}\left(K_{l+1/2}(|\zeta|a)
I_{l+1/2}(|\zeta|a)\right)^n,\ee
and that therefore the total self-energy of the semitransparent sphere is
given by the well-known expression \cite{barton04,Scandurra:1998xa}
\be
E=\frac1{2\pi a}\sum_{l=0}^\infty (2l+1)\int_0^\infty dx\ln\left(1
+\lambda a I_{l+1/2}(x)K_{l+1/2}(x)\right),\quad x=|\zeta|a.
\ee
Actually, a slightly different form involving integration by parts was
given in Refs.~\cite{Milton:2004vy,Milton:2004ya}, 
which results in the energy being finite though order $\lambda^2$.
In order $\lambda^3$ there is a divergence which is associated with
surface energy \cite{CaveroPelaez:2006kq}.

\section{$2+1$ Spatial Geometries} 
We now proceed to apply this method to the interaction between bodies, 
which leads, for example, as Emig et al.\ \cite{Emig:2007cf,Emig:2007me} 
point out, to a multipole expansion.  In this section we illustrate this idea
with a $2+1$ dimensional version, which allows us to describe, for example,
cylinders with parallel axes.  We seek an expansion of the free Green's
function for $\mathbf{R}=\mathbf{R_\perp}$ entirely in the $x$-$y$ plane,
\be
G_0(\mathbf{R+r'-r})=\frac{e^{i|\omega||\mathbf{r-R-r'}|}}{4\pi|
\mathbf{r-R-r'}|}
=\int\frac{dk_z}{2\pi}e^{ik_z(z-z')}g_0(\mathbf{r_\perp-R_\perp-r'_\perp}),
\ee
where the reduced Green's function is
\be
g_0(\mathbf{r_\perp-R_\perp-r'_\perp})=\int\frac{(d^2k_\perp)}{(2\pi)^2}
\frac{e^{-i\mathbf{k_\perp\cdot R_\perp}}e^{i\mathbf{k_\perp\cdot(r_\perp
-r'_\perp)}}}{k_\perp^2+k_z^2+\zeta^2}.\ee
As long as the two potentials do not overlap, so that we have
$\mathbf{r_\perp-R_\perp-r'_\perp}\ne0$, we can write an expansion in
terms of modified Bessel functions:
\be
g_0(\mathbf{r_\perp-R_\perp-r'_\perp})=\sum_{m,m'}I_m(\kappa r) e^{im\phi}
I_{m'}(\kappa r')e^{-im'\phi'}
\tilde g^0_{m,m'}(\kappa R),\quad\kappa^2=k_z^2+\zeta^2.\ee
By Fourier transforming, and using the definition of the Bessel function
\be
i^mJ_m(kr)=\int_0^{2\pi}\frac{d\phi}{2\pi}\,e^{-im\phi}e^{ikr\cos\phi},\ee
we easily find
\be
\tilde g^0_{m,m'}(\kappa R)
=\frac1{2\pi}\int_0^\infty \frac{dk\,k}{k^2+\kappa^2}J_{m-m'}(kR)
\frac{J_m(kr)J_{m'}(kr')}{I_m(\kappa r)I_{m'}(\kappa r')},\ee
which is in fact independent of $r$, $r'$.

As in the previous section, the $k$ integral here can actually be
evaluated as a contour integral, as Bordag noted \cite{Bordag:2006vc}.  
No point-splitting is required here,
because the bodies are non-overlapping, so $r/R, r'/R< 1$.  We write
the dominant Bessel function in terms of Hankel functions,
\be
J_{m-m'}(x)=\frac12\left[H_{m-m'}^{(1)}(x)+H_{m-m'}^{(2)}(x)\right]
=\frac12\left[H_{m-m'}^{(1)}(x)+(-1)^{m-m'+1}H_{m-m'}^{(1)}(-x)\right],
\ee
and then we can carry out the integral over $k$ by closing the
contour in the upper half plane.  We are left with
\be
\int_0^\infty \frac{dx\,x}{x^2+y^2}J_{m-m'}(x)J_m(xr/R)J_{m'}(xr'/R)
=(-1)^{m'}K_{m-m'}(y)I_m(yr/R)I_{m'}(yr'/R),\label{contourint}
\ee
and therefore the reduced Green's function has the simple form
\be
\tilde g_{m,m'}^0(\kappa R)=\frac{(-1)^{m'}}{2\pi}K_{m-m'}(\kappa R).
\ee

Thus we can derive an expression for the interaction 
energy per unit length between two bodies, in
terms of discrete matrices,
\be
\mathfrak{E}\equiv
\frac{E_{\rm int}}{L}=\frac1{8\pi^2}\int d\zeta\,dk_z\ln\det\left(1-\tilde g^0
t_1\tilde g^{0\top} t_2\right),\ee
where $\top$ denotes transpose, and
where the $T$ matrix elements are given by
\be
t_{mm'}=\int dr\,r\,d\phi \int dr'\,r'\,d \phi'I_m(\kappa r)
e^{-im\phi}I_{m'}(\kappa r')e^{im'\phi'}T(r,\phi;r',\phi').\ee

\subsection{Interaction between semitransparent cylinders}
Consider, as an example, two parallel semitransparent cylinders, of
radii $a$ and $b$, respectively, lying
outside each other, described by the potentials
\be
V_1=\lambda_1 \delta(r-a),\quad V_2=\lambda_2\delta(r'-b),
\ee
with the separation between the centers $R$ satisfying $R>a+b$.
It is easy to work out the scattering matrix in this situation,
\be
T_1=V_1-V_1G_0V_1+V_1G_0V_1G_0V_1-\dots,
\ee
so the matrix element is easily seen to be
\be
(t_1)_{mm'}=2\pi\lambda_1a\delta_{mm'}\frac{I_m^2(\kappa a)}{1+\lambda_1 a
I_m(\kappa a)K_m(\kappa a)}.\label{met}
\ee
Again, we used here the regularized integral\footnote{Again, this is
equivalent to the use of the two-dimensional Green's function
$$H_0(kP)=\sum_{m=-\infty}^\infty i^me^{-im\phi'}J_m(k\rho')
e^{im\phi}H^{(1)}_m(k\rho), \quad \rho'<\rho,$$
where $P=\sqrt{\rho^2+\rho^{\prime2}-2\rho\rho'\cos(\phi-\phi')}$.}
\be
\int_0^\infty \frac{dk\,k}{k^2+\kappa^2}J_m(kr)J_m(kr')=K_m(\kappa r')
I_m(\kappa r),\quad r<r'.
\ee
Thus the Casimir energy per unit length is
\be
\mathfrak{E}=\frac1{4\pi}\int_0^\infty d\kappa\,\kappa\,\mbox{tr}\ln(1-A),
\label{eylenergy}
\ee
where
\be
A=B(a)B(b),\ee
in terms of the matrices
\be
B_{mm'}(a)=K_{m+m'}(\kappa R)\frac{\lambda_1 a I_{m'}^2(\kappa a)}
{1+\lambda_1 aI_{m'}(\kappa a)K_{m'}(\kappa a)}.\label{bofa}
\ee
\subsection{Interaction between cylinder and plane}
As a check, let us rederive the result derived by Bordag \cite{Bordag:2006vc}
for a cylinder in front of a Dirichlet plane perpendicular to the $x$ axis.  
We start from the interaction (\ref{gtgt})
written in terms of $\bar G_2$, the deviation from the free Green's function
induced by a single potential,
\be
\bar G_2=G_2-G_0=-G_0T_2G_0,
\ee
so the interaction energy has the form
\be
E=-\frac{i}{2\tau}\mbox{Tr}\ln(1+T_1 \bar G_2).
\ee  When the second body is a Dirichlet plane, $\bar G$ may be found by 
the method of images, with the origin taken at the center of the cylinder,
\be
\bar G(\mathbf{r,r'})=-G_0(\mathbf{r,\bar r'}),\quad \mathbf{\bar r'}=
(R-x',y',z'),
\ee
where $R$ is the distance between the center of the 
 cylinder and its image at $\mathbf{R_\perp}$, that is,
$R/2$ is the distance between the center of the cylinder and the plane.
(We keep $R$ here, rather than $R/2=D$, because of the close connection
to the two cylinder case.)
Now we encounter the 2-dimensional Green's function
\be
g(\mathbf{r_\perp+r'_\perp-R_\perp})=\sum_{mm'}I_m(\kappa r)I_{m'}(\kappa r')
e^{im\phi}e^{im'\phi'}g_{mm'}(\kappa R), \ee
(because the cylinder has $y\to-y$ reflection symmetry)
where the argument given above yields
\be
g_{mm'}(\kappa R)=\frac1{2\pi}K_{m+m'}(\kappa R).
\ee
Thus the interaction between the semitransparent cylinder and a Dirichlet
plane is
\be
\mathfrak{E}=\frac1{4\pi}\int_0^\infty \kappa\,d\kappa\,\mbox{tr}\, 
\ln(1-B(a)),\label{cylpl}\ee
where $B(a)$ is given by Eq.~(\ref{bofa}).
In the strong-coupling limit this result agrees with that given by Bordag,
because
\be
\mbox{tr}\,B^s=\mbox{tr}\,\tilde B^s, 
\quad \tilde B_{mm'}=\frac1{K_m(\kappa a)}K_{m+m'}(
\kappa R) I_{m'}(\kappa a).
\ee

\subsection{Weak-coupling}
\label{sec:wc}
In weak coupling, the formula (\ref{eylenergy})
for the interaction energy between two cylinders
is 
\be
\mathfrak{E}=-\frac{\lambda_1\lambda_2 ab}{4\pi R^2}\sum_{m,m'=-\infty}^\infty
\int_0^\infty dx\,x \,K_{m+m'}^2(x)I_m^2(xa/R)I_{m'}^2(xb/R).\label{wcenergy}
\ee
Similarly, the energy of interaction between a weakly-coupled cylinder
and a Dirichlet plane is from Eq.~(\ref{cylpl})
\be
\mathfrak{E}=-\frac{\lambda a}{4\pi R^2}\sum_{m=-\infty}^\infty \int_0^\infty
dx\,x\ K_{2m}(x)I_m^2(xa/R).\label{wccylpl}\ee

\subsection{Power series expansion}
\label{sec:ps}
It is straightforward to develop a power series expansion for the interaction
between weakly-coupled
semitransparent cylinders.  One merely exploits the small argument
expansion for the modified Bessel functions $I_m(xa/R)$ and $I_{m'}(xb/R)$:
\begin{equation}
I_m^2(x)=\left(\frac{x}{2}\right)^{2|m|}
\sum_{n=0}^\infty Z_{|m|,n} \left(\frac{x}{2}\right)^{2n},
\end{equation}
where the coefficients $Z_{m,n}$ are
\begin{eqnarray}
Z_{m,n} & = & \sum_{k=0}^n \frac{1}{k! \; (n-k)! \;
\Gamma(k+m+1) \; \Gamma(n-k+m+1)}\nn
\\ \label{CoefA} & = & \frac{2^{2(m+n)} \;
\Gamma \! \left( m+n+\frac{1}{2} \right)}{\sqrt{\pi}\;
n! \; (2m+n)! \; \Gamma( m+n+1 ) }.
\end{eqnarray}

The Casimir energy per unit length (\ref{wcenergy}) is now given as
\begin{equation}
\mathfrak{E}= - \frac{\lambda_1 a \lambda_2 b }{4 \pi R^2}
\int_0^\infty  \mathrm{d} x \; x \sum_{m=-\infty}^\infty 
\sum_{m'=-\infty}^\infty\sum_{n=0}^\infty \sum_{n'=0}^\infty
\left(\frac{x a}{2 R}\right)^{2|m|} Z_{|m|,n} \left(\frac{x a}{2 R}\right)^{2n}
\left(\frac{x b}{2 R}\right)^{2|m'|} 
Z_{|m'|,n'}\left(\frac{x b}{2 R}\right)^{2n'} K_{m+m'}^2(x).
\end{equation} 
Reordering terms gives a more compact formula
\begin{equation}
\mathfrak{E}= - \frac{\lambda_1 a \lambda_2 b }{4 \pi R^2}
\sum_{m=-\infty}^\infty \sum_{m'=-\infty}^\infty
\sum_{n=0}^\infty \sum_{n'=0}^\infty
Z_{|m|,n} \left(\frac{a}{R}\right)^{2(|m|+n)}
Z_{|m'|,n'} \left(\frac{b}{R}\right)^{2(|m'|+n')}
J_{|m|+|m'|+n+n',m+m'},
\end{equation}
where the two index symbol $J_{p,q}$ represents
the integral over $x$, which evaluates to
\be
J_{p,q}  =  2 \int_0^\infty dx
\left(\frac{x}{2}\right)^{2p+1} K_{q}^2(x)
=\frac{\sqrt{\pi} \; p! \; \Gamma(p+q+1) \Gamma(p-q+1) }{2^{2p+2}
\Gamma\!\left( p+\frac{3}{2} \right) }.\label{IntegralJ}
\ee

In order to simplify the power series expansion in terms of 
$\tfrac{a}{R}$ and $\tfrac{b}{R}$ we need to reorder the $m$ sums so that only
non-negative values of $m$ appear. 
There are several ways to break up the $m$ sums;
one of them is to decompose the sum into the $m=m'=0$ term, the $m,m'$ same 
sign terms, and the $m,m'$ different sign terms, giving
\begin{multline}
\mathfrak{E}= - \frac{\lambda_1 a \lambda_2 b }{4 \pi R^2}
\left[
\sum_{n=0}^\infty \sum_{n'=0}^\infty
Z_{0,n} \left(\frac{a}{R}\right)^{2n}
Z_{0,n'} \left(\frac{b}{R}\right)^{2n'}
J_{n+n',0}
\right. \\ \left.
+2 \sum_{m=1}^\infty \sum_{m'=0}^\infty
\sum_{n=0}^\infty \sum_{n'=0}^\infty
Z_{m,n} \left(\frac{a}{R}\right)^{2(m+n)}
Z_{m',n'} \left(\frac{b}{R}\right)^{2(m'+n')}
J_{m+m'+n+n',m+m'}
\right. \\ \left.
+2 \sum_{m=0}^\infty \sum_{m'=1}^\infty
\sum_{n=0}^\infty \sum_{n'=0}^\infty
Z_{m,n} \left(\frac{a}{R}\right)^{2(m+n)}
Z_{m',n'} \left(\frac{b}{R}\right)^{2(m'+n')}
J_{m+m'+n+n',m-m'}
\right].
\end{multline}
It is now possible to combine the multiple infinite power series into a single
infinite power series, where each term is given by (possible multiple) finite 
sum(s). In this case we get an amazingly simple result
\be
\mathfrak{E}=-\frac{\lambda_1a\lambda_2b}{4\pi R^2}\frac12\sum_{n=0}^\infty
\left(\frac{a}{R}\right)^{2n}P_n(\mu),\label{multicyl}
\ee
where $\mu=b/a$, and 
where by inspection we identify the binomial coefficients
\be
P_n(\mu)=\sum_{k=0}^n \left(\begin{array}{c}n\\k\end{array}
\right)^2\mu^{2k}.
\ee
Remarkably, it is possible to perform the sums \cite{riordan},
so we obtain the following
closed form for the interaction between two weakly-coupled cylinders:
\be
\mathfrak{E}=-\frac{\lambda_1 a \lambda_2 b}{8 \pi R^2}
\left[\left( 1 - \left( \frac{a+b}{R} \right)^2\right)
\left( 1 - \left( \frac{a-b}{R} \right)^2 \right)\right]^{-1/2}.
\label{excyl}
\ee
We note that in the limit $R-a-b=d\to0$, $d$ being the distance between
the closest points on the two cylinders, we recover the proximity force
theorem in this case (\ref{pfawccyl}),
\be U(d)=-\frac{\lambda_1\lambda_2 }{32\pi}\sqrt{\frac{2ab}R}\frac1{d^{1/2}},
\quad d\ll a, b.
\label{pfawccyl1}
\ee
In Figs.~\ref{figwccyl1}--\ref{figwccyl2}
we compare the exact energy (\ref{excyl}) with the
proximity force approximation (\ref{pfawccyl1}).
Evidently, the former approaches the latter
when the sum of the radii $a+b$ of the cylinders approaches the distance $R$
between their centers. The rate of approach is linear (with slope 3/2)
for the equal radius
case, but with slope $b^2/4a^2$ when $a\ll b$.
More precisely, the ratio of the exact energy to the PFA is
\be
\frac{\mathfrak{E}}U\approx1-\frac{1+\mu+\mu^2}{4\mu}\frac{d}R
\approx 1-\frac{R^2-aR+a^2}{4a(R-a)}\frac{d}R.\label{ratio}
\ee
This correction to the PFA is derived by another method in Appendix \ref{appc}.
The reader should note that the the PFA is actually only defined in the
limit $d\to0$, so the functional form away from that point is ambiguous.
Corrections to the PFA depend upon the specific form assumed for $U(d)$.

\begin{figure}[t]
\vspace{2.in}
\includegraphics{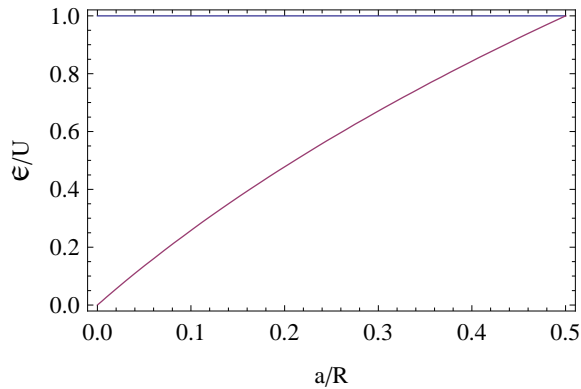}
\caption{\label{figwccyl1}
Plotted is the ratio of the exact interaction energy
(\ref{excyl}) of two weakly-coupled
cylinders to the proximity force approximation
(\ref{pfawccyl1}) as a function of the cylinder radius $a$
for $a=b$.}
\end{figure}

\begin{figure}[t]
\centering
\vspace{2in}
\includegraphics{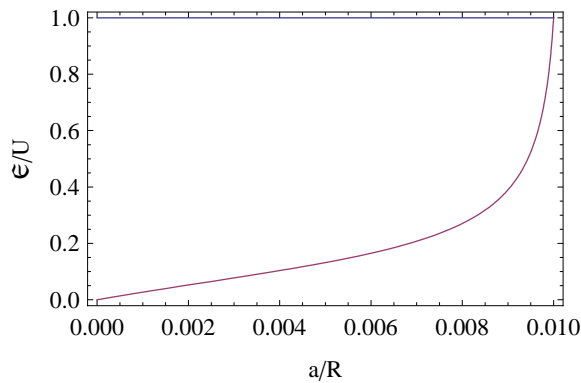}
\caption{\label{figwccyl2}
Plotted is the ratio of the exact interaction energy
(\ref{excyl}) of two weakly-coupled
cylinders to the proximity force approximation
(\ref{pfawccyl1}) as a function of the cylinder radius $a$
for $b/a=99$.}
\end{figure}

\subsection{Exact result for interaction between plane and cylinder}
In exactly the same way, starting from Eq.~(\ref{wccylpl}), we can obtain a
closed-form result for the interaction energy between a Dirichlet plane
and a weakly-coupled cylinder of radius $a$
separated by a distance $R/2$.  The result is again quite
simple:
\be
\mathfrak{E}=-\frac{\lambda a}{4\pi R^2}\left[1-
\left(\frac{2a}R\right)^2\right]^{-3/2}.\label{excylpl}
\ee
In the limit as $d\to0$, this agrees with the PFA:
\be
U(d)=-\frac{\lambda}{64\pi}\frac{\sqrt{2a}}{d^{3/2}}.\label{pfa:cylpl}
\ee
Note again that this form is ambiguous: the proximity force theorem is
equally well satisfied if we replace $a$ by $R/2$, for example, in $U(d)$.
The comparison between this PFA and the exact result (\ref{excylpl})
is given in Fig.~\ref{fig:excylpl}.

\begin{figure}[t]
\vspace{2.in}
\includegraphics{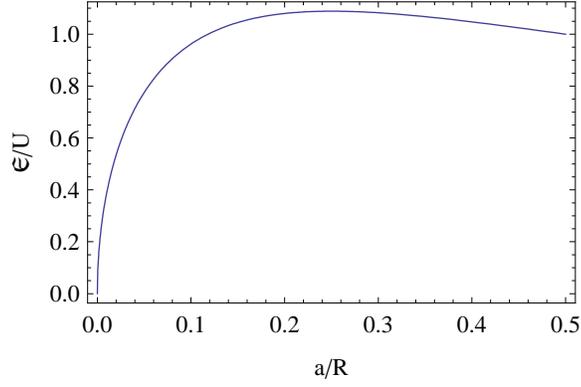}
\caption{\label{fig:excylpl}
Plotted is the ratio of the exact interaction energy
(\ref{excylpl}) of a weakly-coupled
cylinder above a Dirichlet plane to the proximity force approximation
(\ref{pfa:cylpl}) as a function of the cylinder radius $a$.}
\end{figure}

\subsection{Strong coupling (Dirichlet) limit}
The interaction between Dirichlet cylinders 
is given by Eq.~(\ref{eylenergy}) in
the limit $\lambda_1=\lambda_2\to\infty$, that is
\begin{subequations}
\label{scresult}
\be
\mathfrak{E}=\frac1{4\pi R^2}\int_0^\infty dx\,x\,\mbox{tr}\ln(1-A),\label{sccyl}
\ee
where
\be
A_{mm'}=\sum_{m''}K_{m+m''}(x)K_{m''+m'}(x)\frac{I_{m''}(xa/R)}{K_{m''}(xa/R)}
\frac{I_{m'}(xb/R)}{K_{m'}(xb/R)}.
\ee
\end{subequations}
Here the trace of the logarithm can be interpreted as in Eq.~(\ref{logexp}).

Because it no longer appears possible to obtain a closed-form solution, 
we want to verify analytically that as the surfaces of the two cylinders
nearly touch, we recover the result of the proximity force theorem.  
We use a variation of the scheme explained by Bordag for a cylinder
next to a plane \cite{Bordag:2006vc}. [The
analysis is a bit simpler  in the weak-coupling case,
which leads to Eq.~(\ref{pfawccyl1}). See Appendix \ref{appc}.]  First we
replace the products of Bessel functions in $A$ by their leading uniform
asymptotic approximants for all $m$'s large:
\be
B_{mm''}(a)B_{m''m'}(b)
\sim \frac1{2\pi}\frac1{\sqrt{m+m''}}\frac1{\sqrt{m'+m''}}\left(
1+\left(\frac{x}{m+m''}\right)^2\right)^{-1/4}\left(
1+\left(\frac{x}{m'+m''}\right)^2\right)^{-1/4}e^{-\chi},
\ee
where the exponent is
\be
\chi=(m+m'')\eta\left(\frac{x}{m+m''}\right)+
(m'+m'')\eta\left(\frac{x}{m'+m''}\right)-2m''\eta\left(\frac{xa}{m''R}\right)
-2m'\eta\left(\frac{xb}{m'R}\right),
\ee
in terms of
\be
\eta(z)=t^{-1}+\ln\frac{z}{1+t^{-1}},\quad \eta'(z)=\frac1{zt},\quad
\eta''(z)=-\frac{t}{z^2},\label{eta}
\ee
and 
\be
t=(1+z^2)^{-1/2}.
\ee
We write the trace of the $s$th power of $A$ as (summed on repeated indices)
\be
(A^s)_{m_1m_1}=B_{m_1m'_1}(a)B_{m'_1m_2}(b)B_{m_2m'_2}(a)B_{m'_2m_3}(b)
\cdots B_{m_sm'_s}(a)B_{m'_sm_1}(b).
\ee
We rescale variables in terms of a large variable $M$ and relatively small
variables:
\be
m'_i=M\alpha_i,\quad m_i=M\beta_i,
\ee
where without loss of generality we take only $2s-1$ of the $\alpha$'s and
$\beta$'s as independent:
\be
\sum_{i=1}^s(\alpha_i+\beta_i)=s.
\ee
This normalization is chosen so at the critical point where $\chi=0$
for $a+b=R$,
\be
\alpha_i=\frac{a}R \quad \beta_i=1-\frac{a}R, \quad\forall i.
\ee
Away from this point, we consider fluctuations,
\be
\alpha_i=\frac{a}R+\hat \alpha_i,\quad \beta_i=1-\frac{a}R+\hat\beta_i,
\ee
with the constraint
\be
\sum_{i=1}^s(\hat\alpha_i+\hat\beta_i)=0.\label{constraint}
\ee
The Jacobian of this transformation is $sM^{2s-1}$.

Now, we expand the exponent in $\mbox{tr}\,A^s$, to first order in
$d=R-a-b$, and to second order in $\hat\alpha_i$, $\hat\beta_i$.  The result
is
\be
\chi=\frac{2Msd}{tR}+Mt\left(\frac{R}a-1\right)
\sum_{i=1}^s\left[\hat\alpha_i-\frac12\frac{a}{R-a}
(\hat\beta_i+\hat\beta_{i+1})\right]^2
+\frac{Mt}4\frac{a}{R-a}\sum_{i=1}^s(\hat\beta_i-\hat\beta_{i+1})^2.
\ee
The $\hat\alpha_i$ terms lead to trivial Gaussian integrals. The difficulty
with the quadratic $\hat\beta_i$ terms is that only $s-1$ of the differences 
are independent.  But, in view of the constraint (\ref{constraint}) there are
only $s-1$ independent $\beta_i$ variables.  In fact, it is easy to check
that
\be
\sum_{i=1}^s (\hat\beta_i-\hat\beta_{i+1})^2=\sum_{i=1}^{s-1}\frac{i+1}i
\left[\hat\beta_i-\hat\beta_{i+1}+\frac1{i+1}\sum_{j=i+1}^{s-1}(\hat\beta_j-\hat
\beta_{j+1})\right]^2,
\ee
which now enables us to perform each successive $\hat\beta_i-\hat\beta_{i+1}$
integration. The Jacobian of the transformation to the
difference variables $u_i=\hat\beta_i-\hat\beta_{i+1}$, $i=1,\dots,s-1$, is
$1/s$. Thus, we can immediately write down 
\bea
\mathfrak{E}&\sim&-\frac1{4\pi R^2}\int_0^\infty\!\! dz\,z\sum_{s=1}^\infty 
\frac{t^s}s
\int_0^\infty\!\! dM\,\frac{M^{2s+1}}{(2\pi M)^s}e^{-2Msd/tR}
\left[\int_{-\infty}^\infty \!\!\!d\alpha_ie^{-Mt(R-a)\alpha_i^2/a}\right]^s
\prod_{i=i}^{s-1}\int_{-\infty}^\infty du_ie^{-Mat\sum_{i=1}^{s-1}\frac{i+1}i
u_i^2/4(R-a)}\nn\\
&=&-\frac1{4\pi R^2}\int_0^\infty dz\,z\sum_{s=1}^\infty \frac1s\int_0^\infty
dM\,M^{2s+1}\frac{t^s}{(2\pi M)^s}e^{-2Msd/tR}\left[\frac{\pi a}{(R-a)Mt}
\right]^{s/2}\left[\frac{4\pi(R-a)}{Mta}\right]^{(s-1)/2}s^{-1/2}\nn\\
&=&-\frac{\sqrt{2a(R-a)}\pi^3}{3840 R^3}\left(\frac{R}{d}\right)^{5/2},
\eea
which is exactly the result expected from the proximity force theorem, 
according to Eq.~(\ref{scpfa}).

%\subsubsection{Numerical Results}
%A power-series expansion does not exist for the strong-coupling interaction
%between cylinders because $K(z)$ is not analytic at $z=0$.  
%We can, however, extract numerical results from Eq.~(\ref{sccyl}), by 
%truncation of the matrices.  Representative
%results are shown in Fig.~\ref{figs1}--\ref{figs2}, and are compared
%to the proximity force approximation.
%These results are comparable to those given in Refs.~\cite{Emig:2007me}.
%\begin{figure}
%\vspace{2.5in}
%\special{psfile=MultipleScatPlots/MultScatCylStrong2.eps
% hoffset=100 voffset=-0 hscale=90 vscale=90}
%\caption{\label{figs1}
%Truncation of the strong-coupling result (\ref{scresult}) 
%as a function of cylinder radius $a=b$ obtained by truncating the matrices
%to order $(2M+1)\times(2M+1)$ for low values of $M$, $M=1,2,3$, respectively.}
%\end{figure}

%\begin{figure}
%\vspace{2.5in}
%\special{psfile=MultipleScatPlots/MultScatCylStrong1.eps
% hoffset=100 voffset=-0 hscale=60 vscale=60}
%\caption{\label{figs2}
%Truncation of the strong-coupling result (\ref{scresult}) 
%as a function of cylinder radius $a=b$ obtained by truncating the matrices
%to order $(2M+1)\times(2M+1)$ for low values of $M$, $M=1,2,3$, respectively.
%Here the result is given relative to the proximity force approximation.}
%\end{figure}

We will forego further discussion of strong coupling, and presentation of
numerical results, for these have been extensively discussed in several
recent papers, especially in Ref.~\cite{Emig:2007me}.

\section{3-dimensional formalism}
The three-dimensional formalism is very similar.  In this case, the
free Green's function has the representation
\be
G_0(\mathbf{R+r'-r})=\sum_{lm,l'm'}j_l(i|\zeta|r)j_{l'}(i|\zeta|r')
Y_{lm}^*(\mathbf{\hat r})
Y_{l'm'}(\mathbf{\hat r'})g_{lm,l'm'}(\mathbf{R}).\ee
The reduced Green's function can be written in the form
\be
g^0_{lm,l'm'}(\mathbf{R})
=(4\pi)^2i^{l'-l}\int\frac{(d\mathbf{k})}
{(2\pi)^3}\frac{e^{i\mathbf{k\cdot R}}}{k^2+\zeta^2}\frac{j_l(kr)j_{l'}(kr')}
{j_l(i|\zeta| r)j_{l'}(i|\zeta|r')}
Y_{lm}(\mathbf{\hat k})Y^*_{l'm'}(\mathbf{\hat k}).
\ee
Now we use the plane-wave expansion (\ref{pwepw}) once again, this time for
$e^{i\mathbf{k\cdot R}}$,
%\be
%e^{i\mathbf{k\cdot R}}=4\pi\sum_{l''m''}i^{l''}j_{l''}(kR)
%Y_{l''m''}(\mathbf{\hat R})Y_{l''m''}^*(\mathbf{\hat k}),
%\ee
so now we encounter something new, an integral over three spherical
harmonics,
\be
\int d\mathbf{\hat k}Y_{lm}(\mathbf{\hat k})Y_{l'm'}^*(\mathbf{\hat k})
Y_{l''m''}^*(\mathbf{\hat k})=C_{lm,l'm',l''m''},\ee
where
\be
C_{lm,l'm',l''m''}=(-1)^{m'+m''}\sqrt{\frac{(2l+1)(2l'+1)(2l''+1)}{4\pi}}
\left(\begin{array}{ccc}
l&l'&l''\\
0&0&0\end{array}\right)
\left(\begin{array}{ccc}
l&l'&l''\\
m&m'&m''\end{array}\right).
\ee
The three-$j$ symbols (Wigner coefficients) 
here vanish unless $l+l'+l''$ is even.  This fact
is crucial, since because of it we can follow the previous method of
writing $j_{l''}(kR)$ in terms of Hankel functions of the first and second
kind, using the reflection property of the latter,
\be
h_{l''}^{(2)}(kR)=(-1)^{l''}h^{(1)}_{l''}(-kR),
\ee
and then extending the $k$ integral over the entire real axis to a 
contour integral closed in the upper half plane.  The residue theorem
then supplies the result for the reduced Green's function\footnote{This differs
by a (conventional) factor of $|\zeta|$ from the quantity $U_{lml'm'}$ defined
by Emig et al.~\cite{Emig:2007me}.}
\be
g^0_{lm,l'm'}(\mathbf{R})=4\pi  i^{l'-l}\sqrt{\frac{2|\zeta|}{\pi R}}
\sum_{l''m''}C_{lm,l'm',l''m''}K_{l''+1/2}(|\zeta|R)
Y_{l''m''}(\mathbf{\hat R}).
\ee

\subsection{Casimir interaction between semitransparent spheres}
For the case of two semitransparent spheres that are totally outside
each other,
\be
V_1(r)=\lambda_1\delta(r-a),\quad V_2(r')=\lambda_2\delta(r'-b),
\ee
in terms of spherical coordinates centered on each sphere, it is again
very easy to calculate the scattering matrices,
\be
T_1(\mathbf{r,r'})=\frac{\lambda_1}{a^2}\delta(r-a)\delta(r'-a)
\sum_{lm}\frac{Y_{lm}(\mathbf{\hat r})Y_{lm}^*(\mathbf{\hat r'})}{1+\lambda_1a
K_{l+1/2}(|\zeta|a)I_{l+1/2}(|\zeta|a)},\ee
and then the harmonic transform is very similar to that seen in 
Eq.~(\ref{met}), ($k=i|\zeta|$)
\bea
(t_1)_{lm,l'm'}&=&\int (d\mathbf{r})(d\mathbf{r'})
j_l(kr)Y^*_{lm}(\mathbf{\hat r})
j_{l'}(kr')Y_{l'm'}(\mathbf{\hat r'})T_1(\mathbf{r,r'})\nn\\
&=&\delta_{ll'}\delta_{mm'}(-1)^l\frac{\lambda_1 a\pi}{2|\zeta|}
\frac{I_{l+1/2}^2(|\zeta|a)}{1+\lambda_1 a K_{l+1/2}(|\zeta|a)
I_{l+1/2}(|\zeta|a)}.\eea
Let us suppose that the two spheres lie along the $z$-axis, that is, 
$\mathbf{R}=
R\mathbf{\hat z}$.  Then we can simplify the expression for the energy
somewhat by using $Y_{lm}(\theta=0)=\delta_{m0}\sqrt{(2l+1)/4\pi}$.
The formula for the energy of interaction becomes
\be
E=\frac1{2\pi}\int_0^\infty d\zeta \,\mbox{tr}\ln(1-A),\label{etrace}
\ee
where the matrix
\be
A_{lm,l'm'}=\delta_{m,m'}\sum_{l''}B_{ll''m}(a)B_{l''l'm}(b)
\ee
is given in terms of the quantities
\be
B_{ll'm}(a)=\frac{\sqrt{\pi}}{\sqrt{2\zeta R}} i^{-l+l'} \sqrt{(2l+1)(2l'+1)}
\sum_{l''}(2l''+1)\left(\begin{array}{ccc}l&l'&l''\\0&0&0\end{array}
\right)\left(\begin{array}{ccc}l&l'&l''\\m&-m&0\end{array}\right)
\frac{K_{l''+1/2}(\zeta R)\lambda_1 a I_{l'+1/2}^2(\zeta a)}
{1+\lambda_1 a I_{l'+1/2}(\zeta a)K_{l'+1/2}(\zeta a)}.
\ee
Note that the phase always cancels in the trace in Eq.~(\ref{etrace}).
For strong coupling, this result reduces to that found by Bulgac, Wirzba
et al.~\cite{Bulgac:2005ku,wirzba07} for Dirichlet spheres, and recently
generalized by Emig et al.~\cite{Emig:2007me} 
for Robin boundary conditions.  (See also Ref.~\cite{hensler}.)

\subsection{Weak coupling}
For weak coupling, a major simplification results because of the
orthogonality property,
\be
\sum_{m=-l}^l\left(\begin{array}{ccc}l&l'&l''\\m&-m&0\end{array}
\right)\left(\begin{array}{ccc}l&l'&l'''\\m&-m&0\end{array}\right)
=\delta_{l''l'''}\frac1{2l''+1},\quad l\le l'.
\ee
Then the formula for the energy of interaction between the two spheres is
\be
E=-\frac{\lambda_1a\lambda_2b}{4R}\int_0^\infty \frac{dx}x\sum_{ll'l''}
(2l+1)(2l'+1)(2l''+1)\left(\begin{array}{ccc}l&l'&l''\\0&0&0\end{array}
\right)^2
K_{l''+1/2}^2(x)I_{l+1/2}^2(xa/R)I_{l'+1/2}^2(xb/R).
\ee
There is no infrared divergence because for small $x$ the product of Bessel
functions goes like $x^{2(l+l'-l'')+1}$, and $l''\le l+l'$.

As with the cylinders, we expand the modified Bessel functions
of the first kind in power series in $a/R,b/R<1$. This expansion yields
the infinite series
%\begin{multline}
%E= - \frac{\lambda_1 a \lambda_2 b }{4 R} 
%\int_0^\infty \frac{\mathrm{d} x}{x} 
%\sum_{l=0}^\infty\sum_{n=0}^\infty\sum_{l'=0}^\infty
%\sum_{n'=0}^\infty\sum_{l''=0}^\infty
%(2l+1)(2l'+1)(2l''+1)\begin{pmatrix}l & l' & l'' \\ 0 & 0 & 0 \end{pmatrix}^2
%\\ \times
%\left(\frac{x a}{2 R}\right)^{2l+1} A_{l+\frac{1}{2},n} 
%\left(\frac{x a}{2 R}\right)^{2n}
%\left(\frac{x b}{2 R}\right)^{2l'+1} A_{l'+\frac{1}{2},n'} 
%\left(\frac{x b}{2 R}\right)^{2n'}
%K_{l''+\frac{1}{2}}^2(x),
%\end{multline}
%still using Eq.~\eqref{CoefA} for the definition of $A_{m,n}$. 
%Slightly reordering the terms, and carrying out the integration we get
%\begin{multline}
%E= - \frac{\lambda_1 a \lambda_2 b }{32 R} \; \frac{a b}{R^2}
%\sum_{l=0}^\infty\sum_{n=0}^\infty\sum_{l'=0}^\infty
%\sum_{n'=0}^\infty
%\left(\frac{a}{R}\right)^{2(l+n)}
%\left(\frac{b}{R}\right)^{2(l'+n')} 
%\\
%\times \sum_{l''}
%(2l+1)(2l'+1)(2l''+1)\begin{pmatrix}l & l' & l'' \\ 0 & 0 & 0 \end{pmatrix}^2
%A_{l+\frac{1}{2},n} A_{l'+\frac{1}{2},n'}
%J_{l+n+l'+n',l''+\frac{1}{2}},
%\end{multline}
%where $J_{p,q}$ is still given by Eq.~\eqref{IntegralJ}. 
%The $l''$ sum is carried out over the
%region dictated by the triangle property of the Wigner 3-$j$ symbol, 
%which states that  $|l-l'| \le l'' \le l+l'$. 
\begin{equation}
E=- \frac{\lambda_1 a \lambda_2 b }{4 \pi R}
\frac{a b}{R^2}
\sum_{n=0}^\infty \frac{1}{n+1} \sum_{m=0}^n
D_{n,m}\left(\frac{a}{R}\right)^{2(n-m)}\left(\frac{b}{R}\right)^{2m},
\label{mesphere}
\end{equation}
%where the $C_{n,m}$ coefficients are given by
%\begin{multline}
%C_{n_1,n_2}=\frac\pi8(n_1+1)\sum_{n_3=0}^{n_1-n_2} \sum_{n_4=0}^{n_1}
%\sum_{l''}
%(2(n_1-n_2-n_3)+1) (2(n_2-n_4)+1) (2l''+1)
%\\ 
%\times\begin{pmatrix} (n_1-n_2-n_3) & (n_2-n_4) & l'' \\0&0&0\end{pmatrix}^2
%A_{n_1-n_2-n_3+\frac{1}{2},n_3}A_{n_2-n_4+\frac{1}{2},n_4}
%J_{n_1,l''+\frac{1}{2} },
%\end{multline}
%where the $l''$ sum is again deliminated by the triangle
%property of the Wigner 3-$j$ symbols. 
where by inspection of the first several $D_{n,m}$ coefficients
we can identify them as
\begin{equation}
D_{n,m}=\frac{1}{2} \binomial{2n+2}{2m+1},
\end{equation}
and now we can immediately sum the expression (\ref{mesphere})
 for the Casimir interaction energy to give 
the closed form 
\begin{equation}
E=\frac{\lambda_1 a \lambda_2 b}{16 \pi R} \; \ln\left(
\frac{1-\left(\frac{a+b}{R}\right)^2}{1-
\left(\frac{a-b}{R}\right)^2}\right).\label{exsphere}
\end{equation}

Again, when $d=R-a-b\ll a,b$, the proximity force theorem (\ref{pfawcsphere}) 
is reproduced:
\be
U(d)\sim \frac{\lambda_1\lambda_2ab}{16\pi R}\ln (d/R),\quad
d\ll a, b.\label{pfawcsphere1}
\ee  
However, as Figs.~\ref{fig3}, \ref{fig4} demonstrate, the
approach is not very smooth,
even for equal-sized spheres. The ratio of the energy to the PFA is
\be
\frac{E}{U}=1+\frac{\ln[(1+\mu)^2/2\mu]}{\ln d/R},\quad d\ll a, b,
\ee
for $b/a=\mu$. Truncating the power series (\ref{mesphere}) at $n=100$
would only begin to show the approach to the proximity force theorem limit.  
The error in using the PFA between spheres can be very substantial.

\begin{figure}[t]

\vspace{2in}
\includegraphics{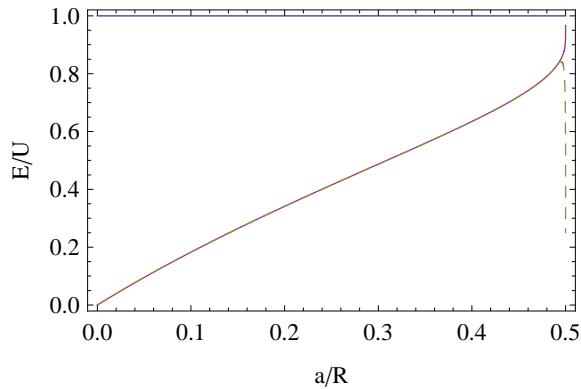}
\caption{\label{fig3}
Plotted is the ratio of the exact interaction energy
(\ref{exsphere}) of two weakly-coupled
spheres to the proximity force approximation
(\ref{pfawcsphere1}) as a function of the sphere radius $a$
for $a=b$.  Shown also by a dashed line is the power series expansion
(\ref{mesphere}), truncated at
$n=100$, indicating that it is necessary to include
very high powers to capture the proximity force limit.}
\end{figure}

\begin{figure}[t]

\vspace{2in}
\includegraphics{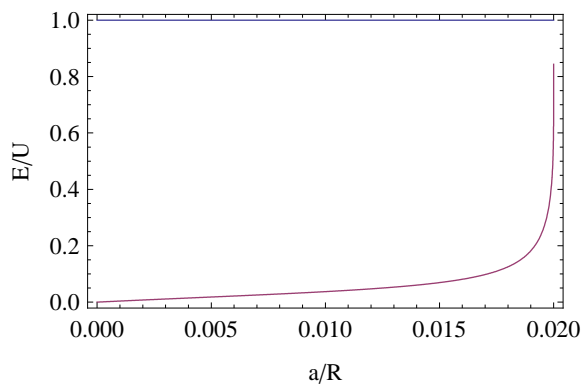}
\caption{\label{fig4}
Plotted is the ratio of the exact interaction energy
(\ref{exsphere}) of two weakly-coupled
spheres to the proximity force approximation
(\ref{pfawcsphere1}) as a function of the sphere radius $a$
for $b/a=49$.}
\end{figure}

Again we will forego discussion of the strong-coupling (Dirichlet) limit
here because of the extensive discussion already in the literature
\cite{Bulgac:2005ku,wirzba07,Emig:2007me}.
 
\subsection{Exact result for interaction between plane and sphere}
In just the way indicated above, we can obtain a closed-form result
for the interaction energy between a weakly-coupled sphere and a Dirichlet
plane.  Using the simplification that
\be
\sum_{m=-l}^{l} (-1)^m 
\left(\begin{array}{ccc}l&l&l'\\m&-m&0\end{array}\right)
\left(\begin{array}{ccc}l&l&l'\\0&0&0\end{array}\right)
=\delta_{l'0},
\ee
we can write the interaction energy
in the form
\be
E=-\frac{\lambda a}{2 \pi R} \int_{0}^{\infty}
dx \sum_{l=0}^{\infty}
\sqrt{\frac{\pi}{2x}} (2l+1) K_{1/2}(x)
I_{l+1/2}^2(x(a/R)).
\ee
Then in terms of $R/2$ as the distance between the center of the sphere and the
plane, the exact interaction energy is
\be
E=-\frac\lambda{2\pi}\left(\frac{a}R\right)^2\frac1{1-(2a/R)^2},\label{exsppl}
\ee
which as $a\to R/2$ reproduces the proximity force limit, contained in the
(ambiguously defined) PFA formula
\be
U=-\frac\lambda{8\pi}\frac{a}{d}.\label{pfa:sppl}
\ee
The exact energy and this PFA approximation are compared in 
Fig.~\ref{fig:exsppl}.

\begin{figure}[t]

\vspace{2in}
\includegraphics{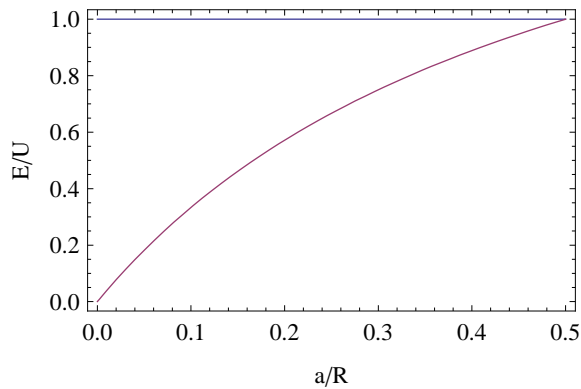}
\caption{\label{fig:exsppl}
Plotted is the ratio of the exact interaction energy
(\ref{exsppl}) of a weakly-coupled
sphere above a Dirichlet plane to the proximity force approximation
(\ref{pfa:sppl}) as a function of the sphere radius $a$.}
\end{figure}

\section{Comments and Prognosis}
%The methods proposed recently are in fact not particularly novel, being
%utilized in this context in 
Although the multiple scattering methods date
back to the 1970s \cite{renne,Balian:1977qr}
%What is new is the ability, largely due to enhancement in computing
%power and flexibility, to evaluate continuum determinants (or infinitely
%dimensional discrete ones) accurately numerically. 
%This is making it possible to compute Casimir forces for geometries
%previously inaccessible. 
some remarkable new results have been obtained.
 Here we have given a perhaps simpler and more
transparent derivation of the procedure than in Refs.~\cite{Emig:2007cf,
Emig:2007me}.  For example, because we have approached the problem from a
general field theoretic viewpoint, we see that 
the ``translation matrix'' introduced there
is nothing other than the free Green's function.  
Our approach yields the general form first, and the multipole
expansion as a derived consequence, not the other way around.
We apply this multiple scattering
method to obtain new results for the interaction between semitransparent
cylinders and spheres, and we have analytically demonstrated the approach to 
the proximity force theorem.  Most remarkably, we have derived explicit,
very simple, closed-form expressions for the interaction between weakly
coupled cylinders and between weakly coupled spheres,
as well as between weakly-coupled cylinders or spheres and
Dirichlet planes.  These explicit
results demonstrate the profound limitation of the proximity force 
approximation, which has been under serious attack for some
time \cite{reynaud1,reynaud2}. We hope that
these developments will lead to improved conceptual understanding, and 
to better comparison with experiment, when they are extended to realistic 
materials.

\begin{acknowledgments}
We thank the US National Science Foundation (Grant No.\ PHY-0554926) and the
US Department of Energy (Grant No.\ DE-FG02-04ER41305) for partially funding
this research.  We thank Prachi Parashar and K. V. Shajesh for extensive
collaborative assistance throughout this project.
  We are grateful to many participants, and particularly to
 the organizer Michael Bordag, in the workshop
on Quantum Field Theory Under the Influence of External Conditions held
in Leipzig in September 2007 (QFEXT07) for many illuminating lectures and
discussions.  We are appreciative of Steve Fulling's suggestion that we
investigate this subject.
\end{acknowledgments}

\appendix
\section{Derivation of Vacuum Energy Formula}
\label{Appa}
Following Schwinger \cite{Schwinger75}
we start from the vacuum amplitude in terms of sources,
\be
\langle 0_+|0_-\rangle^K=e^{iW[K]},\quad
W[K]=\frac12\int(dx)(dx')K(x)G(x,x')K(x').\label{a1}\ee
Here $G$ is the Green's function in the presence of some background potential.
From this the effective field is
\be
\phi(x)=\int(dx')G(x,x')K(x').\ee
If the geometry of the region is altered slightly, as through moving one
of the bounding surfaces, the vacuum amplitude is altered:
\be
\delta W[K]=\frac12\int (dx)(dx')K(x)\delta G(x,x')K(x')
=-\frac12\int (dx)(dx')\phi(x)\delta G^{-1}(x,x')\phi(x'),\label{a3}\ee
which uses the fact that
\be
G G^{-1}=1.\ee
Upon comparison of Eq.~(\ref{a3}) with the two particle emission term in
\be
e^{iW[K]}=e^{i\int(dx)K(x)\phi(x)+i\int(dx)\mathcal{L}}
=\dots+\frac12\left[i\int (dx)K(x)\phi(x)
\right]^2,
\ee
we deduce from the coefficient of $\phi(x)\phi(x')$ 
that the effective two-particle source due to a geometry
modification is
\be
iK(x)K(x')\bigg|_{\rm eff}=-\delta G^{-1}(x,x').\label{a6}\ee
Thus the change in the generating functional is obtained by inserting
Eq.~(\ref{a6}) into Eq.~(\ref{a1}),
\be
\delta W=\frac{i}2\int (dx)(dx')G(x,x')\delta G^{-1}(x',x)
=-\frac{i}2\int(dx)(dx')\delta G(x,x') G^{-1}(x',x),
\ee
which gives the change in the action under an alteration of the Green's
function. From this, in matrix notation
\be
\delta W=-\frac{i}2\delta\mbox{Tr}\ln G\Rightarrow E=\frac {i}{2\tau}
\mbox{Tr}\ln G,\label{a8}\ee
because for a static configuration $W=-E\tau$,
which is our starting point, Eq.~(\ref{trln}).

There are of course many other derivations of this famous result.
For example, one can derive it rather simply on the basis of Schwinger's
quantum action principle.  It may also be worth noting that it is formally
equivalent to another familiar representation for the quantum vacuum energy
\be 
E= -i\int_{-\infty}^\infty \frac{d\omega}{2\pi}\omega^2\mbox{Tr}\,\mathcal{G},
\label{o2g}\ee
(for example, see Ref.~\cite{Milton:2004ya}).  Here, the Fourier transform
of the Green's function appears,
\be
G(x,x')=\int_{-\infty}^\infty \frac{d\omega}{2\pi}e^{-i\omega(t-t')}
\mathcal{G}(\mathbf{r,r'};\omega).
\ee
In terms of $\mathcal{G}$, Eq.~(\ref{a8}) can be written in the form
\be
E=\frac{i}2\int_{-\infty}^\infty \frac{d\omega}{2\pi}\mbox{Tr}\,\ln\mathcal{G}.
\label{trlnG}\ee
Because 
\be
\mathcal{G}^{-1}\mathcal{G}=1,\quad \mathcal{G}^{-1}=-\omega^2-\nabla^2+V,
\ee
we see that when Eq.~(\ref{trlnG}) is integrated by parts, and
surface terms are ignored, we immediately recover Eq.~(\ref{o2g}).

\section{Proximity Force Approximation}
\label{appb}
In this Appendix we derive the proximity force approximation (PFA) for the
energy of interaction between two semitransparent cylinders, or two 
semitransparent spheres, either in the strong or weak coupling regimes.
This approximation, relating the force between nonplanar surfaces in terms
of the forces between parallel plane surfaces, was first introduced in
1934 by Derjaguin \cite{derjaguin}.
This approximation is only valid when the separation between the bodies
is very small compared to their sizes.  It is now well established that
the approximation cannot be extended beyond that limit, and that
1\% errors occur if the PFA is applied when the ratio of the separation
to the radius of curvature of the bodies is of order 1\%.  Fortunately,
current experiments have not exceeded this limit.  That should change
in the near future, which is one reason the new numerical calculations
are of importance.  In fact, we have found that in general the errors
in using the PFA may be much larger than indicated above.
We concur with Bordag that while the proximity force theorem is exact
at zero separation, any approximation based on extrapolation away from that
point is subject to uncontrollable errors.

\begin{figure}
\vspace{2.5in}
\includegraphics{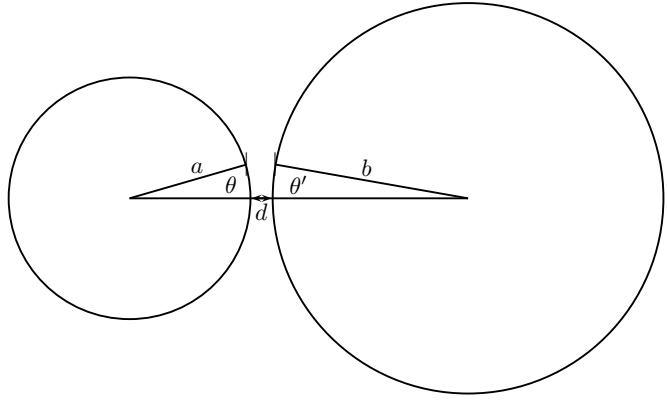}
\caption{Geometry of two cylinders (or two spheres) with radii $a$ and
$b$, respectively, and distances between their centers of $R>a+b$.
The proximity force approximation applies when the distance of closest
approach $d=R-a-b\ll a,b$.  The approximation consists in assuming
that the interaction is dominated by the interaction of adjacent surface
elements, as shown.}
\label{fig1}
\end{figure}

Consider first two parallel cylinders, of radius $a$ and $b$, with their
centers separated by a distance $R>a+b$.  The distance of closest approach
of the cylinders is $d=R-a-b$.  The PFA consists of assuming that
the energy between the two bodies is the sum of the energies between small
parallel plane elements at the same height along the surfaces, that is,
in polar coordinates at $\theta$ relative to the center of cylinder a
and at $\theta'$ relative to the center of cylinder b, where
as seen in Fig.~\ref{fig1},
\be
a\sin\theta=b\sin\theta'.
\ee  
Because $d$ is much smaller than either $a$ or $b$, only small values of
$\theta$ actually contribute, and the energy of interaction $U(d)$ between the
surface may be expressed in terms of the energy per area $\mathcal{E}(h)$
for the corresponding parallel plate problem, with separation distance $h$:
\be
U(d)=\int a\,d\theta\, \mathcal{E}[d+a(1-\cos\theta)+b(1-\cos\theta')].
\ee
Here, for weak coupling [see Eq.~(\ref{wcpp})],
\be
\mathcal{E}(h)=-\frac{\lambda_1\lambda_2}{32\pi^2 h}.
\ee
Because $\theta$ is small, the PFA energy per length is
\be
U(d)=-\frac{\lambda_1\lambda_2}{32\pi^2}
\frac{a}{d}\int_{-\pi}^\pi d\theta\left[1+\frac{a}d\left(1+\frac{a}b
\right)\frac{\theta^2}2\right]^{-1}
=-\frac{\lambda_1\lambda_2 }{32\pi}\sqrt{\frac{2ab}R}\frac1{d^{1/2}}.
\label{pfawccyl}
\ee
To obtain the corresponding result for strong coupling, we merely
replace $\mathcal{E}(h)=-\pi^2/(1440 h^3)$, and a similar calculation yields
\be
U(d)=-\frac{\pi^3}{3840}\sqrt{\frac{2ab}R}\frac1{d^{5/2}},\quad
d\ll a, b.\label{scpfa}
\ee

It is easy to reproduce the result given by Bordag \cite{Bordag:2006vc}
for a cylinder in front
of a plane.  For the strong coupling (Dirichlet) case we simply take 
the result (\ref{scpfa}) and regard $b$ as much larger than $a$, and
obtain
\be
U(d)=-\frac{\pi^3}{1920\sqrt{2}}\frac{a^{1/2}}{d^{5/2}},\quad d\ll a.
\ee
For a weakly coupled cylinder in front of a Dirichlet plane, we start
from the corresponding interaction between two such planes, $\mathcal{E}(h)
=-\lambda/(32\pi^2h^2)$, which leads to
\be
U(d)=-\frac\lambda{64\pi}\frac{(2a)^{1/2}}{d^{3/2}}.
\ee

For nearly touching spheres the calculation goes just the same way.
The result, for strong coupling (Dirichlet boundary conditions), 
for the PFA energy is
\be
U(d)=-\frac{\pi^3}{1440}\frac{ab}R\frac1{d^2},\quad d\ll a, b,
\ee
while in the weak-coupling limit there is sensitivity to large $\theta$
signifying a logarithmic divergence,
\be
U(d)\sim \frac{\lambda_1\lambda_2ab}{16\pi R}\ln (d/R),\quad
d\ll a, b.\label{pfawcsphere}
\ee
For a weakly-coupled sphere in front of a Dirichlet plane,
a PFA approximation is
\be
U(d)=-\frac\lambda{16\pi}\frac{a}{d}.
\ee

\section{Short distance limit}
\label{appc}
\subsection{Cylinders}
In this appendix we want to discuss the short distance limit, for
the case of weakly-coupled cylinders, where 
the closest distance between the cylinders is $R-a-b=d\ll a, b$,
which should reduce to the proximity force approximation derived in 
Appendix \ref{appb}.  We will calculate the first correction to the PFA,
and compare to the exact result found in Sec.~\ref{sec:ps}.
In this limit, we replace the modified Bessel functions
by their uniform asymptotic approximants, which in leading form yield
\be
K_{m+m'}^2(x)I_m^2(xa/R)I_{m'}^2(xb/R)\sim \frac1{8\pi}\frac1{mm'(m+m')}
tt_at_b e^{-\chi},
\ee
where
\be
t=(1+z^2)^{-1/2}, \quad t_a=(1+z_a^2)^{-1/2},\quad t_b=(1+z_b^2)^{-1/2},
\ee
and
\be
z=\frac{x}{m+m'},\quad z_a=\frac{xa/R}{m},\quad z_b=\frac{xb/R}{m'}.\ee
The exponent here is
\be
\chi=2(m+m')\eta(z)-2m\eta(z_a)-2m'\eta(z_b),\label{chi}
\ee
where $\eta$ is defined by Eq.~(\ref{eta}).
The reason that the force diverges as $a+b\to R$ is that $\chi$ vanishes
there, for suitable values of $m$ and $m'$.  To make this
systematic, let us
rescale variables,
\be
m=M\alpha,\quad m'=M\beta,
\ee
and then when $b=R-a$, $\chi=0$ when $\beta a=\alpha b$.

When $b=R-a-d$, with $d$ small compared to the radius of either cylinder,
we assume that the main contribution comes from the neighborhood of these
values.  So we define
\be
\alpha=\frac{a}R+\hat\alpha,\quad \beta=1-\frac{a}R+\hat\beta, 
\ee
and we expand the exponent to first order in $d$ and to second order in 
$\hat\alpha$ and $\hat\beta=-\hat \alpha$. (The latter constraint ensures
that $\alpha+\beta=1$.) The result is
\be
\chi=\frac{2Md}{tR}+\frac{MtR^2\hat\alpha^2}{a(R-a)}+O(\hat\alpha^3,d^2).
\ee
Then
\bea
\mathfrak{E}
&\sim&-\frac{\lambda_1\lambda_2}{16\pi^2}\int_0^\infty dz\,z\,t^3\int_0^\infty
dM \,e^{-2Md/tR}\int_{-\infty}^\infty 
d\hat\alpha\, e^{-Mt\hat\alpha^2R^2/[a(R-a)]}\nn\\
&=&-\frac{\lambda_1\lambda_2}{32\pi}\sqrt{\frac2d}\sqrt{\frac{a(R-a)}R}=U,
\label{pfaform1}
\eea
which is exactly the result given by the proximity force theorem in Appendix
\ref{appb}, Eq.~(\ref{pfawccyl}).

Now we calculate the correction to the PFA.  We do this by keeping subleading
terms in the uniform asymptotic approximation for the product of six
Bessel function
\bea
K_{m+m'}^2(x)I_m^2(xa/R)I_{m'}^2(xb/R)&\sim&\frac{1}{8\pi m m'}\frac{tt_at_b}{
m+m'}\nn\\
&&\quad\times\left(1-\frac{u_1(t)}{m+m'}\right)^2
\left(1+\frac{u_1(t_a)}{m}\right)^2\left(1+\frac{u_1(t_b)}{m'}\right)^2
e^{-\chi},
\eea
where $t=t(z)$ with $z=x/(m+m')$, $z_a=xa/m$, $z_b=xb/m'$,
\be
u_1(t)=\frac{3t-5t^3}{24},
\ee
and $\chi$ is given by Eq.~(\ref{chi}).  Now when we expand $\chi$ we
must go out to order $\hat\alpha^4$, $d^2$, and $\hat\alpha^2 d$.  The result
is
\bea
e^{-\chi}&\sim& e^{-2Md/tR}e^{-M\hat\alpha^2 tR^2/(a(R-a))}\bigg[1-\frac{d^2Mt}
{R(R-a)}+\frac{2\hat\alpha d Mt}{R-a}\nn\\
&&\quad\mbox{}-\frac{\hat\alpha^2 d M t(1-t^2)R}{(R-a)^2}
+\frac{\hat\alpha^3 M t^3(R-2a)R^3}{3a^2(R-a)^2}+\frac{\hat\alpha^4
Mt^3(1-3t^2)(R^2-3aR+3a^2)R^4}{12 a^3(R-a)^3}\nn\\
&&\quad\mbox{}+\frac{2M^2\hat\alpha^2d^2 t^2}{(R-a)^2}+\frac{M^2\hat\alpha^6t^6
(R-2a)^2R^6}{18a^4(R-a)^4}+\frac23\frac{M^2\hat\alpha^4 t^4dR^3}{a^2(R-a)^3}
(R-2a)\bigg].
\eea
As above, we replace 
\be
m=M\frac{a}R\left(1+\hat\alpha \frac{R}a\right),\quad 
m'=M\left(1-\frac{a}R\right)\left(1-\hat\alpha \frac{R}{R-a}\right).
\ee
We expand $t_a$ and $t_b$ in the prefactor using
\be
\frac{dt}{dz}=-zt^3,\quad \frac{d^2t}{dz^2}=2t^3-3t^5.
\ee
The PFA is obtained by using the integrals
\begin{subequations}
\bea
\int_{-\infty}^\infty d\hat\alpha\,e^{-\hat\alpha^2\gamma}&=&
\sqrt{\frac{\pi}{\gamma}},
\quad \gamma=MtR^2/a(R-a),
\\
\int_0^\infty \frac{dM}{\sqrt{M}}e^{-2Md/t}&=&\Gamma\left(\frac12\right)\left(
\frac{2d}t\right)^{-1/2},
\eea
\end{subequations}
and so, from the expansion we can obtain the result of the integrals
over $\hat \alpha$ and $M$ by the algebraic substitutions
\begin{subequations}
\bea
\frac1M&\to& -\frac{4d}{Rt},\quad M\to \frac{tR}{4d},\\
\hat\alpha^2&\to&-\frac{2a(R-a)d}{R^3t^2},\quad M\hat\alpha^2\to\frac12
\frac{a(R-a)}{R^2t},\quad M\hat\alpha^4\to-\frac{3a^2(R-a)^2d}{R^5t^3},\\
M^2\hat\alpha^2&\to&\frac{a(R-a)}{8Rd},\quad
M^2\hat\alpha^4\to \frac34\frac{a^2(r-a)^2}{R^4t^2},\quad
M^2\hat\alpha^6\to-\frac{15}2\frac{a^3(R-a)^3d}{R^7t^4}.
\eea
\end{subequations}
The result is the following correction factor to the PFA in the
form given in Eq.~(\ref{pfaform1}):
\be
\frac{\mathfrak{E}}U
=1-\frac{R^2+aR+a^2}{4a(R-a)}\frac{d}R.
\ee
Although this looks slightly different from Eq.~(\ref{ratio}), it agrees
with the latter when the  PFA formula (\ref{pfawccyl1}) 
is expressed in terms of the form given in Eq.~(\ref{pfaform1}), that is,
writing $d=R-a-b$.

\subsection{Spheres}

Here we see how the proximity force limit is achieved
for weakly-coupled spheres.  Again, the strategy
is to replace the modified Bessel functions by their leading uniform
asymptotic approximants.  The only new element is the appearance of the
$3$-$j$ symbol.  Because now only $m=0$ appears, there is a very simple
approximant for the latter \cite{ponzano,biedenharn,aquilanti}:
\be
\left(\begin{array}{ccc}l&l'&l''\\0&0&0\end{array}\right)\sim\sqrt{\frac\pi2}
\frac{\cos\frac\pi2(l+l'+l'')}{[(l+l'+l'')(l+l'-l'')(l-l'+l'')
(-l+l'+l'')]^{1/4}}.
\ee
This result is quite accurate, being within 1\% of the true value
of the Wigner coefficient for $l$'s of order 100 (except very near the
boundaries of the triangular region, where the approximant diverges weakly).
Otherwise, the procedure is rather routine.  Letting $\nu=l+1/2$, and similarly
for the primed quantities, we expand the exponent resulting from the 
uniform asymptotic expansion about the critical point, with
\be
\nu=N(a+\hat\alpha),\quad \nu'=N(1-a+\hat \alpha'),\quad \nu''=N(1+\alpha''),
\ee
with the constraint $\hat\alpha+\hat\alpha'+\hat\alpha''=0$.  Replacing the
sums over angular momenta by integrals, and changing variables:
\be
\int dv\,d\nu'\,d\nu''=2\int_0^\infty dN\,N^2 \int_0^\infty d(\hat\alpha+\hat
\alpha')\int_{-\infty}^\infty \frac{d(\hat\alpha-\hat\alpha')}2,
\ee
which reflects the restriction emerging from the triangular relation of
the Wigner coefficients, $\hat\alpha+\hat \alpha'>0$, we find for the
approximant to the energy when the two spheres are nearly touching:
\bea
E&\sim&-\frac{\lambda_1a\lambda_2b}{4R}\frac2\pi\int_0^\infty \frac{dx}x
\int_0^\infty dN\,N^2 t^3 e^{-2Nd/Rt}\frac1{4\pi N^2}\left[\frac{R^2}{a(1-a)}
\right]^{1/2}
\int_0^\infty \frac{d(\hat\alpha+\hat\alpha')}{(\hat\alpha+\hat\alpha')^{1/2}}
e^{-4N(\hat\alpha+\hat\alpha')/t}\nn\\
&&\qquad\times\int_{-\infty}^\infty d\left(\frac{\hat\alpha
-\hat\alpha'}2\right)e^{-NtR^2(\hat\alpha-\hat\alpha')^2/4a(R-a)}\nn\\
&\sim&\frac{\lambda_1\lambda_2 ab}{16\pi R}\ln d,\quad d=R-a-b\ll a, b,
\eea
which is exactly the PFA result (\ref{pfawcsphere}).

\end{document}